\documentclass[preprint,journal]{IEEEtran} \usepackage[dvipsnames, usenames, table]{xcolor} 
\usepackage{hyperref}
\usepackage{amsmath,amsfonts}
\usepackage{algorithmic}
\usepackage{algorithm}
\usepackage{array}
\usepackage{graphicx}
\usepackage{tabularx}
\usepackage{pifont}

\usepackage{booktabs}

\usepackage{textcomp}
\usepackage{stfloats}
\usepackage{url}
\usepackage{verbatim}
\usepackage{graphicx}
\usepackage{cite}
\usepackage{wrapfig}
\usepackage{arydshln} %
\usepackage{colortbl} 
\usepackage{tabu}                      %
\usepackage{mwe}                       %
\usepackage{csquotes}
\usepackage{multirow}

\usepackage{mathptmx}      

\usepackage{makecell}

\newcommand{\rev}[1]{\textcolor{black}{#1}}

\makeatletter
\newcommand{\thickhline}{%
    \noalign {\ifnum 0=`}\fi \hrule height 1pt
    \futurelet \reserved@a \@xhline
}

\begin{document}

\title{\rev{Cognitive Affordances in Visualization}:\\ Related Constructs, Design Factors, and Framework}

\author{Racquel Fygenson, Lace Padilla, \textit{Member, IEEE}, and Enrico Bertini 
\thanks{All authors are with Northeastern University. \\ \indent E-mail: fygenson.r $\vert$ l.padilla $\vert$ e.bertini@northeastern.edu}%
\thanks{Manuscript received Month Day, Year; revised Month Day, Year.}}

\markboth{Journal of \LaTeX\ Class Files,~Vol.~14, No.~8, August~2021}%
{Shell \MakeLowercase{\textit{et al.}}: A Sample Article Using IEEEtran.cls for IEEE Journals}

\maketitle
\begin{abstract}
Classically, affordance research investigates how the shape of objects communicates actions to potential users. \rev{Cognitive affordances, a subset of this research, characterize how the design of objects influences cognitive actions, such as information processing.} Within visualization, \rev{cognitive} affordances inform how graphs' design decisions communicate information to their readers. Although  several related concepts exist in visualization, a formal translation of affordance theory to visualization is still lacking.
In this paper, we review and translate affordance theory to visualization \rev{by formalizing how cognitive affordances operate within a visualization context}. We also review common methods and terms, and compare related constructs to \rev{cognitive} affordances \rev{in visualization}. Based on a synthesis of research from psychology, human-computer interaction, and visualization, we propose a framework of \rev{cognitive} affordances in visualization that enumerates design decisions and reader characteristics that influence a visualization’s hierarchy of \rev{communicated information}. Finally, we demonstrate how this framework can guide the evaluation and redesign of visualizations.
\end{abstract}


\section{Introduction}
\label{sec:intro}
\IEEEPARstart{T}{he} data visualization community has a long history of evaluating how visualizations function and can be improved. We often examine how design decisions, such as shapes, colors, and formats, impact graphs and the ways readers use them.
Our community has developed a plethora of evaluations to assess visualizations, which can be broadly grouped into \textit{performance}\rev{~\cite{lam-bertini-2012}} and \textit{affective}\rev{~\cite{lee-robbins-affective-vis}} evaluations. \textbf{\textit{Performance}} evaluations examine how well visualization users can extract information, and evaluate accuracy and response time \cite{lam-bertini-2012, cleveland-mcgill-1987, quadri-perception-survey, heer-bostock-repr, maceachren-semiotics, fischer-bars-orientation}, and memorability\cite{borkin-memorability}.
\textbf{\textit{Affective}} evaluations investigate how visualizations impact readers' feelings, opinions, or values \cite{lee-robbins-affective-vis} (e.g., perceptions of risk \cite{padilla-covid-risk}, empathy  \cite{boy-bertini-anthropomorphism}, trust \cite{padilla-mfvs-trust}). These evaluations are essential, but do not fully capture an important aspect of visualizations: the information that is most likely communicated. This information includes both knowledge that is explicitly displayed and understanding that can be inferred, such as insights, patterns, and relationships between variables. This information and its likelihood of being communicated is made possible by the strategic creation, manipulation, and arrangement of visual elements (i.e., \textbf{design decisions}), and can be impacted by the background and abilities of different readers (i.e., \textbf{reader characteristics}). \rev{In this paper, we formalize the relationship between visualization design decisions, reader characteristics, and the likelihood of conveying information using a framework of \textbf{cognitive affordances}, which we demonstrate can be used to reason about design implications and create guidelines for more intentional visualizations.}

\begin{wrapfigure}{l}{0.25\textwidth} %
    \centering
    \includegraphics[width=0.25\textwidth, alt={Left: A bar chart with two bars labeled A and B. Right: a line chart with dots at the same height of the bar chart's bars, also labeled A and B.}]{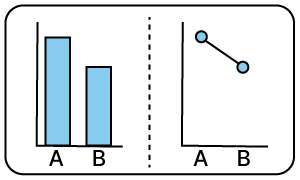}
    \vspace{-4mm}
    \caption{These two charts encode identical data, but \rev{the bar chart affords discrete categories, whereas the line chart communicates a single continuous variable}~\cite{zacks-tversky-bars-lines, shah-graph-comprehension-1999}.}
    \label{fig:linevbar}
    \vspace{-2mm}
\end{wrapfigure}

In contrast to physical affordances, which describe how objects' designs influence users' actions\cite{norman-psych-of-everyday-things, norman-1999}, \textbf{\textit{cognitive affordances}} describe the relationship between the design of an object and the knowledge that \rev{is imparted upon} the object's user \cite{Hartson-2003}. \rev{Investigating visualizations' cognitive affordances} can reveal the likelihood or strength of communicated information. For example, Figure \ref{fig:linevbar} shows two identical datasets, each graphed with a different \textit{encoding mark} (i.e., visual object that represents the data~\cite{bertin1967semiologie}). Both charts convey that A is larger than B, but the left chart's bars suggest a discrete relationship between A and B, whereas the right chart's line implies a continuous relationship, perhaps due to lived experience with visually similar physical objects\cite{zacks-tversky-bars-lines, shah-bar-line-graph-comprehension}. Although performance or affective evaluations of these charts may show differences in outcomes (e.g., actions, judgments, emotional responses) they fail to explain how visualizations convey information in ways that drive these differences. Focusing solely on outcomes leaves us without a detailed theoretical understanding of how visualizations influence performance and affect, limiting progress in the field.

\rev{Cognitive} affordances can help us create more purposeful graphs by revealing \textit{how} and \textit{why} design decisions impact readers' perception, probable use, and understanding of visualized information. \rev{As we demonstrate in Section \ref{sec:comparing-design-decisions}}, visualization\rev{s' cognitive} affordances can link designs to misconstrued information and help designers create visualizations that are more aligned with their intentions. %

Although \rev{cognitive} affordances are not commonly referenced in visualization, we are not the first to consider visualizations' differences in communication. \rev{Visualization} researchers have explored concepts related to \rev{cognitive} affordances, using distinct terms such as ``intuitiveness,'' ``naturalness,'' ``ease,'' and ``spontaneous interpretation'' to describe similar constructs. %
In this paper, we provide an overview of these disparate terms and propose a framework of \rev{cognitive} affordances \rev{in visualization} to offer structure for connecting existing contributions \rev{in visualization} and create a more systematic foundation for future research.

To further visualization research using affordance theory, we contribute:
\begin{enumerate}
    \item a \textbf{translation of affordance theory} to visualization
    \item  an %
    \rev{\textbf{overview}} of affordance-related terms in visualization
    \item a synthesis of themes into a \rev{cognitive-affordance} \textbf{framework} \rev{for visualization}
    \item a demonstration of how the framework can \textbf{heuristically evaluate visualizations}, \textbf{generate predictions}, and \textbf{suggest alternative designs}.
\end{enumerate}

In Section \ref{sec:background-aff}, we review a history of affordance theory, and \rev{describe how cognitive affordances manifest within visualization}. In Section \ref{sec:existing_research}, we \rev{contextualize cognitive affordances} within past visualization research.
In Section \ref{sec:theory}, we propose a framework of \rev{cognitive} affordances \rev{in visualization} and in Section \ref{sec:framework-application}, we illustrate practical applications of the framework. Finally, in Sections \ref{sec:discussion} \rev{and \ref{sec:future-work-limitations}},
we outline open \rev{visualization} research questions surrounding \rev{cognitive} affordances and detail limitations of the current framework.

\section{Background on Affordances}
\label{sec:background-aff}
\subsection{Affordance in Psychology}
``Affordance'' was first coined by ecological psychologist J.J. Gibson in 1979 to describe all actions that an environment can offer or support animals to take~\cite{gibson-ecological-vis-perception}. For example, Gibson characterized surfaces that are fairly flat, horizontal, extended, and rigid (all in comparison to an animal's weight, size, and ability) as affording support to that animal.
Gibson emphasized that affordances depend on actors' perception, as unperceived affordances cannot influence actions~\cite{gibson-ecological-vis-perception, Kaptelinin_aff_encyclo}. For example, an animal will not stand on a surface it does not perceive as stand-able, regardless of the surface's actual affordance. At the same time, Gibson described affordances as system properties created by the interaction of actors and environments, neither purely objective nor subjective~\cite{gibson-ecological-vis-perception, zhang-dist-cog-aff}.

B\ae{}rentsen and Trettvik~\cite{baerentsen-activity-theory-2002} argue that Gibson's definition of affordances as relationships between objects and actors rejects the classical Stimulus-Response theory in psychology (e.g., \cite{hollan2008cognitive}), which views subjects' actions as a direct response to their environment. They propose that Gibson's affordance is more in line with Activity theory, which views actions as the culmination of an actor's cultural and contextual background in addition to their environment \cite{baerentsen-activity-theory-2002}. Motivated by an Activity theory approach to affordances,
B\ae{}rentsen and Trettvik advocate for increased consideration of individual differences for effective use of affordance theory in design practice.

In the decades since Gibson introduced affordances, researchers have expanded the concept to inform other psychological frameworks, physical and digital interaction design, and sociological theories \cite{gaver-aff-1991, zhang-dist-cog-aff, Norman_revised, norman-design-of-everyday-things, davis2016socio}. For example, Gaver posits that affordances can be not only visual, but also tactile, and can build upon each other in a nested or sequential arrangement \cite{gaver-aff-1991}, \rev{and Hartson et al. propose that affordances can manifest in not only physical, but also cognitive, actions \cite{hartson-1999-user-actionframework}}. Gaver also differentiates between ``perceptible," ``hidden," and ``false" affordances, incorporating perception as a defining construct of affordances \cite{gaver-aff-1991}. Zhang and Patel apply a distributed cognition lens to affordances, also emphasizing perception by positing that an environment can afford by not only supporting actions, but also ``inviting" them \cite{zhang-dist-cog-aff}. Zhang and Patel also %
reinforce the concept of cognitive affordances originally proposed by Hartson et al. as actions that are made known by learned convention (e.g., stop signs afford stopping to licensed drivers) \cite{zhang-dist-cog-aff, hartson-1999-user-actionframework}. %

\subsection{Affordance in Human-Computer Interaction (HCI)}
In 1988, Norman adopted and popularized using affordances in design\cite{norman-psych-of-everyday-things}. In contrast to Gibson's definition of affordance, Norman, although perhaps not intentionally~\cite{norman-1999}, communicated objects' affordances as the total set of \textit{perceived} possible actions that an actor could perform with that object \cite{norman-psych-of-everyday-things, Kaptelinin_aff_encyclo}.

For example, a designer might describe a button that is not obviously clickable as ``not affording clicking", even if the button can be physically clicked. Newer work by Norman argues that actions that are made obvious by an object's form are not affordances, but rather ``signifiers"~\cite{Norman_revised}, although this distinction has not permeated design terminology~\cite{Kaptelinin_aff_encyclo}.

An inherent binary is commonly understood to apply to affordances in HCI (e.g., a button either affords or does not afford clicking).  McGrenere and Ho expand upon Norman's work~\cite{norman-psych-of-everyday-things}, arguing that affordances can occur to various degrees (e.g., a button can strongly or weakly afford clicking)~\cite{mcgrenere-ho-aff-2000}. In support of a theory of variable affordances, Franchak and Adolph present psychophysical data to suggest that affordances can be mapped to probabilistic functions~\cite{franchak-adolph-aff-2014}.

 HCI affordance research largely focuses on physical actions, such as clicking or moving a mouse, or selecting an option from a dropdown. %
However, visualizations do more than allow for physical interactions; their primary use is to communicate information. Effectively designed graphs also afford their readers the task of conceptualizing the graphs' underlying information. \rev{This paper establishes} the affordance of mental actions (e.g., extracting, comparing, and understanding information) \rev{as \textbf{\textit{cognitive affordances}}}, which are fundamental to understanding how visualizations communicate. 

\subsection{Cognitive Affordances in Visualization}
We adopt Norman's \rev{intended definition} of affordances \rev{as the total set of possible actions that an actor can perform with an object, modifying it with} three key distinctions supported by Hartson's \cite{hartson-1999-user-actionframework, Hartson-2003}, B\ae{}rentsen and Trettvik's \cite{baerentsen-activity-theory-2002}, and McGrenere and Ho's \cite{mcgrenere-ho-aff-2000} work to define a \rev{cognitive} affordance framework \rev{in visualization}. \rev{Firstly, although Norman's definition of affordance is tied} to physical actions, \rev{we use precedent from Hartson and Zhang and Patel's work to assert that \textbf{affordances}} \textbf{can also encompass mental actions}~\cite{hartson-1999-user-actionframework, Hartson-2003, zhang-dist-cog-aff}. Although visualizations can lead to multiple types of affordances, our \rev{framework} describes cognitive affordances in visualization\footnote{Physical affordances in visualization pertain to interactivity with charts and can be informed by other HCI affordance work. See Section \ref{sec:discussion} for discussion.}.

Our second \rev{modifier of Norman affordances} is that \rev{all} affordances in visualization, \rev{including cognitive affordances,} \textbf{exist as a relationship between a visualization's design and its reader.} %
This distinction aligns with Activity theory \cite{activity-theory-chapter} and has a basis in B\ae{}rentsen and Trettvik's \rev{affordance} work~\cite{baerentsen-activity-theory-2002} that emphasizes object design and actor traits. Our framework (see Sec. \ref{sec:theory}) indicates that \rev{cognitive} affordances \rev{in visualization}, and their varying strengths, result from the interaction of design choices and reader characteristics.

Our \rev{third distinction from Norman affordances} is that \textbf{a hierarchy often exists within a set of \rev{cognitive} affordances.} That is to say, a visualization can afford any information that may be gleaned from it, but may afford some pieces of information more strongly than others. This comparative strength has a basis in work from McGrenere and Ho \cite{mcgrenere-ho-aff-2000}, and Franchak and Adolph~\cite{franchak-adolph-aff-2014} that reject a binary characterization of affordances. In Section \ref{sec:affordance-right}, we discuss past visualization findings that provide evidence of this hierarchical nature.

\section{Related Research in Visualization}
\label{sec:existing_research}
Past \rev{visualization} research has used multiple terms for concepts related to \rev{cognitive} affordance\rev{s}. \rev{Below, we provide an overview of related concepts in visualization to contextualize the contributions of this paper, and we encourage future work to complete a comprehensive literature review to fully map the landscape. To craft this overview,} we identified \rev{six} seminal works that are highly cited ($>$300 citations) and pertain to our research focus: how does a visualization design imply information to its readers? (~\cite{zacks-tversky-bars-lines, friel-graph-sense, shah-graph-comprehension-1999, mackinlay-auto-design, maceachren-semiotics, vessey-cog-fit}).
\rev{From this foundation, we examined both works referenced within these articles and subsequent works citing them within visualization, informatics, and psychology journals. We screened in stages: first, reviewing papers' titles for words that align with our focus and next examining abstracts to determine whether they addressed our methodological or conceptual interests. Then, we limited paper inclusion by degree of theoretical characterization; we excluded papers that introduced only constructs colloquially without theoretical definition or relation to a broader theoretical framework. We read the remaining papers in full and conducted the same citation searches on them in an iterative process.} We continued this review process until our sampling no longer rendered new terms (i.e., reached thematic saturation)\rev{, yielding 21 papers. In Section \ref{sec:rel-constr}, we define and compare these papers' constructs to cognitive affordances. While we did not snowball sample off the constructs that failed to meet our theoretical inclusion criteria, we still mention these papers' constructs in a collective subsection, as their findings are relevant to our original research question.}

\subsection{Related Constructs}
\label{sec:rel-constr}
Numerous publications propose \rev{cognitive} affordance-related constructs within visualization (see terms and definitions in Table \ref{tab:terms}). 
 \rev{We provide an overview of related constructs to better contextualize our framework in current research.} %
\newline

\subsubsection{\textbf{Perceived Visual Salience}}
Visual salience examines the elements on which a visualization reader focuses and is commonly studied by eye-tracking~\cite{Kourtis-affordance-salience-objects, poole-eye-tracking}. In contrast, \textit{perceived} visual salience is assessed through self-report measures, wherein participants identify visualization elements they perceive as most prominent~\cite{xiong-curse-knowledge}.

\textit{Relation to  \rev{Cognitive} Affordance:} Visual elements that are highly salient may contribute strongly to the information that a visualization affords. For example, spacing a group of visual objects close to one another, and far from other objects, can make them more salient, which in turn can increase the strength with which this grouping is communicated. %
Salience can drive attention modulation, whereas \rev{cognitive} affordances can result from it~\cite{Kourtis-affordance-salience-objects, carswell-spontaneous-interpretation}. 
\vspace{-4mm}
\begin{table}[h!]
\caption{\rev{Cognitive} Affordance-Related Constructs \rev{in Visualization}}
\small 
\begin{tabularx}{\columnwidth}{>{\raggedright\arraybackslash}p{2.5cm}
>{\raggedright\arraybackslash}X
>{\raggedright\arraybackslash}p{1cm}}
\toprule
\textbf{Related Construct} & \textbf{Definition \& References} \\
\midrule
Self-Reported \linebreak Visual Salience & The most noticeable and important feature(s) \cite{xiong-curse-knowledge} \\
\hdashline
Intuitiveness & ``extent to which symbols are directly apprehended or readily understood" \cite{maceachren-semiotics, golbiowska-color-intuitiveness, hegarty-cartography-intuitions} \\
\hdashline
Perceived \linebreak Difficulty & \rev{subjective judgment of the ease of completing a task}~\cite{culbertson-1959-graph-comprehension, golbiowska-color-intuitiveness }\\
\hdashline
Cognitive Fit & The degree of alignment between the visual representation of information and the viewer's mental schemas~\cite{vessey-cog-fit} \\
\hdashline
Cognitive\linebreak Naturalness \& Encoding-Message Correspondence &   The intuitive way that visual representations align
with our ingrained spatial experiences and understanding of the world \cite{zacks-tversky-bars-lines, tversky-spatial-schema} \\

\hdashline
Expressiveness & ``A set of facts is expressible in a language if it contains a sentence that (1) encodes all the facts in the set, (2) encodes only the facts in the set." \cite{mackinlay-auto-design, mackinaly-expressiveness-language}, \\
\hdashline
Congruence \linebreak Principle & ``The content and format of [a] graphic should correspond to the content and format of the concepts to be conveyed" \cite{tversky-animation-facilitate}\\
\hdashline
Graph Comprehension & ``Relatively effortless and automatic retrieval of some quantitative concepts, alongside the effortful and complex induction of other quantitative concepts" ~\cite{shah-bar-line-graph-comprehension, friel-graph-sense, shah-graph-comprehension-1999, Shah2002_graph_comprehension_book, Fox2023_graph_comprehension, pinker-graph-comp} \\
\hdashline
\bottomrule
\label{tab:terms}
\end{tabularx}
\vspace{-8mm}
\end{table}

\subsubsection{\textbf{Intuitiveness}}
\label{sec:intuitiveness}
Intuitiveness or intuitions in visualization has been described as the ``extent to which symbols are directly apprehended or readily understood" \cite{maceachren-semiotics}. However, some researchers opt not to define the terms, instead relying on its colloquial use\cite{golbiowska-color-intuitiveness, hegarty-cartography-intuitions}. Within the broader scope of HCI, intuitiveness has been equated to a system's design leading to effective (i.e., ``adequate, exact, and complete") interaction, without providing users with additional knowledge \cite{Naumann-intuitive-use-HCI}. %
This definition highlights accuracy and efficiency in evaluating intuitiveness, but we caution against equating intuitiveness with effective use, as interfaces and visualizations can foster incorrect intuitions.
We discuss this in Section \ref{sec:unintentional} on ``unintentional affordances.''

\textit{Relation to \rev{Cognitive} Affordances:} Intuitiveness emphasizes the importance of aligning visualization design with desired communication goals and is a key motivator for further understanding \rev{cognitive} affordance\rev{s}. Intuitiveness explores the representational fit of a visualization for its underlying data (i.e., ``does design D communicate concept C?''), whereas \rev{cognitive} affordances investigate the strengths with which underlying information is readily communicated to its audience (i.e., ``what does design D communicate first? second? third?''). 
\newline
\subsubsection{\textbf{Perceived Difficulty}}
Some visualizations aim to optimize effort, assuming that \rev{charts} perceived as easier to understand are more desirable~\cite{quadri-perception-survey,culbertson-1959-graph-comprehension, golbiowska-color-intuitiveness}. 
Other work argues that increased cognitive effort in visualizations can promote deeper understanding, challenging the notion that easier is always better~\cite{hullman-visual-difficulties}. Although cognitive effort is commonly measured by proxies such as readers' response time when completing tasks or eye tracking, some work asks readers to report perceived effort (i.e., how easy or difficult they feel it is to use a visualization to complete a task)~\cite{golbiowska-color-intuitiveness, culbertson-1959-graph-comprehension}. 

\textit{Relation to \rev{Cognitive} Affordances:} Logically, information that is easier to perceive can be more likely to be communicated, and vice versa. It may be possible, however, for a visualization to modulate reader attention such that one piece of information is strongly emphasized above others, regardless of ease. Future work investigating the extent of \rev{the two metrics'} correlation is warranted. We provide concrete examples and explain the relationship between perceived difficulty and \rev{cognitive} affordance further in Section \ref{sec:ease}.
\newline 
  \subsubsection{\textbf{Cognitive Fit}}
  \label{sec:cog-fit}
Vessey's Cognitive Fit theory explores the alignment between visual representations and readers' \textit{mental schemas} (i.e., organization of graphic conventions that are stored in the mind)\cite{vessey-cog-fit}.  
    Cognitive Fit theory proposes that visualizations that closely align with readers' mental schemas require less mental transformation, and therefore are more efficient and less prone to incorrect interpretation \cite{vessey-cog-fit-empirical-study, vessey-cog-fit}.
    
    \textit{Relation to \rev{Cognitive} Affordances:} %
    Cognitive Fit theory suggests that viewers activate mental schemas to interpret visualizations, with errors arising from mismatches between schemas and design. This concept aligns with the underlying principles of cognitive affordance, particularly that visualization designs prompt users to interpret and load relevant pieces of encoded information. Unlike Cognitive Fit, which relies on reader performance as a proxy for schema alignment, \rev{cognitive} affordances directly examine how design choices elicit specific information, enabling finer-grained analysis without proxies. \rev{Whereas a visualization's cognitive fit might inform how quickly or correctly readers can use the graph to complete tasks, cognitive affordances would inform which tasks readers are most likely to complete unprompted.}
\newline
\subsubsection{\textbf{Cognitive Naturalness and Encoding-Message Correspondence}}
\label{sec:em-correspondence}
Highly cited related work compares bar charts to line charts and introduces the concept of \textit{cognitive naturalness} (i.e.,  the intuitive way that visual representations align with our ingrained spatial experiences and understanding of the world)~\cite{zacks-tversky-bars-lines}. This concept posits that humans' daily interactions with the physical world shape visualization understanding, as evidenced by universal visual communication practices across different cultures and age ranges.

\textit{Relation to \rev{Cognitive} Affordances:} Cognitive naturalness is exemplified by Gestalt principles, such as grouping or continuity, that illustrate how people naturally organize and interpret visual information\cite{tversky-cognitive-origins}. As \rev{mentioned} in Section \ref{sec:intro}, Zacks and Tversky demonstrate that certain visualizations tend to elicit specific interpretations based on their structural characteristics. Line charts prompt readers to conceptualize data as continuous, whereas bar charts suggest discrete comparisons, a phenomenon that Zacks and Tversky refer to as ``bar-line message correspondence'' \cite{zacks-tversky-bars-lines}. This finding depicts how visual objects' form can initiate cognitive processes and imply information (e.g., continuity or discreteness) that is not extractable from raw numbers, providing strong precedent for \rev{cognitive} affordance in visualization. \rev{Cognitive} affordance, however, goes beyond the encoding analysis of cognitive naturalness to investigate the impact of other design decisions (e.g., data shape, annotations, etc.) on communicated information.
    
Additionally, examinations of cognitive naturalness %
aim to determine a single representation that best corresponds to a dataset, suggesting a one-to-one match between ideal visualization and interpretation. \rev{Cognitive} affordance, on the other hand, suggests a hierarchy where multiple pieces of information can be afforded simultaneously and \rev{considers reader differences that undermine the concept of finding universal optimal visualization-interpretation matches}.
 \newline 
    \subsubsection{\textbf{Expressiveness}}
    Expressiveness is concerned with expressing only and all relevant information \cite{mackinlay-auto-design}. The concept of expressiveness has seen limited development, and there is still no detailed understanding of how specific visualization components express particular types of information. The most advanced treatment of this idea is found in Munzner's textbook~\cite{munzner2015visualization}, which organizes visual channels by their ability to express categorical, quantitative, and ordinal data.
    
    \rev{In introducing expressiveness, Mackinlay contrasts it to ``effectiveness,'' or how well visualizations ``exploit'' human capabilities to extract the information they encode~\cite{mackinlay-auto-design}. %
    As we discuss in Section~\ref{sec:intro}, cognitive affordances are orthogonal to affective and performance visualization measures, and occupy a similar place in their ability to augment expressiveness and effectiveness measures by advising on the likelihood of conveyed information, not just the appropriateness or accuracy of graphic representations' fit.}

\textit{Relation to \rev{Cognitive} Affordances:} 
    Expressiveness \rev{is centered around a visualization design's effect on the entire set of possible extractable information, without regard for differences in the likelihood that readers extract ``expressed'' information. For example, using expressiveness one may conclude that design D expresses concepts A, B, and C, whereas cognitive affordances would help one distinguish that design D is very likely to communicate concept A to Reader R, and less likely to communicate concepts B and C.
    Additionally, %
    in practice,} expressiveness literature mostly focuses only on visual channels' communication, without consideration for other design factors.%
\newline
    \subsubsection{\textbf{Congruence Principle}}
    The \textit{congruence principle} states that ``the content and format of [a] graphic should correspond to the content and format of the concepts to be conveyed''~\cite{tversky-animation-facilitate}. This principle has been applied to specific design decisions, such as the appropriate use of animation\cite{tversky-animation-facilitate}. Similar to \hyperref[sec:em-correspondence]{cognitive naturalness}, the congruence principle advocates that some graphic representations are inherently universally understood and informed by spatial experience. This hypothesis is based on recurring themes of spatial representations throughout human history \cite{tversky-cognitive-origins, tversky-spatial-schema, tversky-animation-facilitate}.

    \textit{Relation to \rev{Cognitive} Affordances:} Although the congruence principle and cognitive affordances \rev{can both be used to identify pain points of graphic designs, the congruence principle is grounded in the goal of minimizing reader confusion (i.e., design D does not convey concept C---design E would better convey concept C), whereas cognitive affordances also inform  understanding of reader interpretation and hierarchy of conveyed information (i.e., design D conveys concept F more strongly than concept C---design E would convey concept C more strongly)}.
\newline
\subsubsection{\textbf{Graph Comprehension}}
\label{sec:graph-comp}
Graph comprehension involves linking numeric data and visual representations to the mental models that viewers develop from them~\cite{shah-bar-line-graph-comprehension, friel-graph-sense, shah-1995-comprehending-line-graphs, Shah2002_graph_comprehension_book, pinker-graph-comp}. Graph comprehension differs from previously described concepts by focusing on both \textit{top-down} and \textit{bottom-up} cognitive processes. Bottom-up cognitive processes involve low-level visual identification, where information is extracted from visual features, and can be described by concepts such as intuitiveness or cognitive naturalness. Conversely, top-down processing incorporates semantic content, long-term knowledge, and expertise when interpreting graphs. Graph comprehension exemplifies how the integration of bottom-up and top-down processes forms a feedback loop, enabling viewers to interpret complex visual representations. For a review of graph comprehension, see \cite{Fox2023_graph_comprehension}. For a discussion of cognitive processing in visualization interpretation, see \cite{Parsons2014-distributed-cognition}

\textit{Relation to \rev{Cognitive} Affordances:} 
The relationship between \rev{cognitive} affordances and graph comprehension is informed by Shah et al.'s observation that graph comprehension involves both effortless, automatic retrieval of some quantitative concepts and effortful, complex induction of others~\cite{Shah2002_graph_comprehension_book}. Prior knowledge and expertise interact with bottom-up processes, such as perceptual effort, resulting in individual differences through top-down processes.
Graph comprehension research highlights that some mappings occur automatically, but it lacks a systematic evaluation of the elements driving this ease. \rev{We can use cognitive} affordances to \rev{systematically evaluate which visual elements correspond to specific types of information retrieval} %
offering a deeper understanding of how the processes \rev{that make up graph comprehension} integrate to \rev{impact viewers' takeaways}.
\newline

\subsubsection{\rev{\textbf{Other Constructs}}}
\label{sec:other-constructs}
\rev{Visual comprehension \cite{quadri-doyouseewhatisee}, comparison and visual affordances \cite{xiong-afford-comparison, xiong-grouping-cues, xiong-thesis}, spontaneous interpretation \cite{carswell-spontaneous-interpretation}, preferences for communication\cite{levy-tversky-gratuitous-graphics}, and statistical communication \cite{acarturk-graphical-cues} also appear to investigate similar effects in visualization as cognitive affordances, and use similar methods as the other constructs we discuss. These five constructs, however, do not have strong theoretical backgrounds (e.g., theory papers that characterize them, definitions contextualized within other cognitive theories) with which we can confidently differentiate them from or equate them to cognitive affordances. Still, due to these five constructs' similar motivations, it is likely that their research can inform our understanding of cognitive affordances in visualization.}

\subsection{Experimental Methods}
\label{sec:methods}
Despite variation in the terminology of affordance-related constructs from the prior section, there is a notable similarity in their methodological evaluation.
 In the following section, we summarize past experimental approaches \rev{in visualization} used to study concepts related to \rev{cognitive} affordances.
\newline

\subsubsection{\textbf{Describe a Graph}}
\label{sec:describe-graph}
The majority of related visualization research employs the ``describe-a-graph method", in which participants answer (either in writing or aloud) open-ended questions about visualizations they are shown~\cite{quadri-doyouseewhatisee, carswell-spontaneous-interpretation, zacks-tversky-bars-lines, shah-1995-comprehending-line-graphs, acarturk-graphical-cues, shah-graph-comprehension-1999, xiong-afford-comparison, shah-bar-line-graph-comprehension, xiong-grouping-cues}. For example, Zacks and Tversky, and Acartürk asked participants to ``describe in a sentence what is shown'' ~\cite{zacks-tversky-bars-lines, acarturk-graphical-cues}. Other works instructed participants to ``describe what [they] see in the graph"\cite{quadri-doyouseewhatisee, carswell-spontaneous-interpretation, shah-graph-comprehension-1999, shah-1995-comprehending-line-graphs}. Further examples include asking participants to describe patterns in bar charts through text and visual annotations~\cite{xiong-afford-comparison}, summarize main points in two to four sentences~\cite{shah-bar-line-graph-comprehension}, or provide their first, second, and third conclusions~\cite{xiong-grouping-cues}.

 After collecting free responses from participants, researchers often \textit{thematically code} results by extracting common themes in responses and synthesizing the data~\cite{alsaawi2014coding}. Importantly, free-response investigations are constrained by the themes that researchers document through their qualitative coding. For example, Shah and Carpenter coded only whether graphed variables were described by participants in a nominal, ordinal, or metric fashion \cite{shah-1995-comprehending-line-graphs}. Similarly, Xiong et al. coded all of their participants' free responses within the context of 12 types of comparison they defined in a ``top-down" approach (i.e., pre-generated prior to coding), leading to their evaluation of specific ``comparison affordances'' \cite{xiong-afford-comparison}.

Describe-a-graph experiments measure the end result of a cognitive process %
and we contend that the results of this process are shaped by visualizations' \rev{cognitive} affordances. %
\newline
\subsubsection{\textbf{Draw a Graph}}
Free-response methods also encompass approaches that show data to participants (e.g., data tables, textual descriptions of data) and ask them to create a corresponding visualization. Although less common due to the challenge of qualitatively coding visuals, these methods explore similar \rev{phenomena} as describe-a-graph approaches. For example, Zacks and Tversky used free-response drawing to re-test their describe-a-graph experiments comparing bar and line charts~\cite{zacks-tversky-bars-lines}.
\newline
\begin{figure*}[t!]
    \centering
\includegraphics[alt={A large box on the far left is labeled Design Decisions and lists five subcategories: Encoding, Arrangemnet, Contextualizing Visual Elements, Data, and Situation. This box flows into an elongated oval in the middle of the figure labels Readers. In the Readers oval there are four subcategories: Demographics, Prior Beliefs, Learned Skills, and lastly Physical \& Cognitive Function. The flowing continues from the Design Decisions box, into the Readers oval, and then into the last box on the far right, titled Afforded Information. The afforded information box shows many different texts reading Info A, Info R, Info G, etc. at different heights of the box. The Info texts that are higher up are a darker green to correlate with an arrow that runs most of the height of the box and is labeled ``More Afforded'' at the top and ``less afforded'' at the bottom. Along the bottom of the box, under a dotted line, are some Info texts in gray labeled ''Not Afforded''}, width=0.9\textwidth]{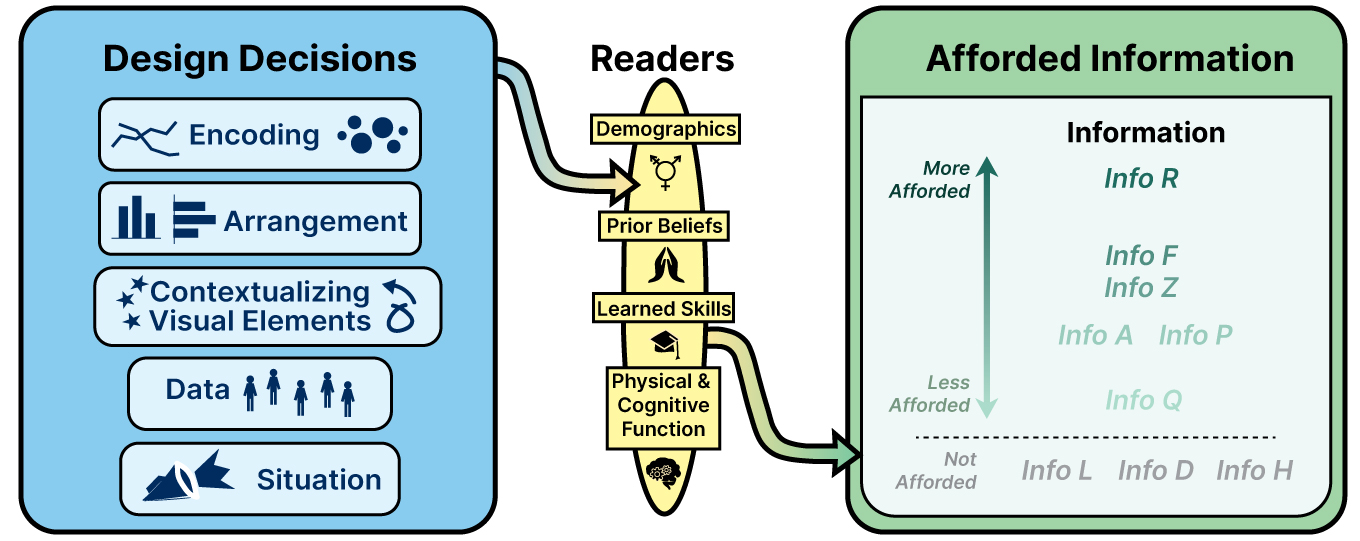}
    \vspace{-2mm}
    \caption{Our framework of cognitive affordances \rev{in visualization consists of} three high-level components: design decisions, readers, and a hierarchy of afforded information. The impacts of design decisions are moderated by reader characteristics, \rev{which dictate the likelihood that different information is communicated}.}
    \label{fig:teaser}
    \vspace{-3.5mm}
\end{figure*}

\subsubsection{\textbf{Select a Graph}}
 ``Select-a-graph" methods \rev{use alternative forced choice tasks~\cite{alt-forced-choice},} presenting participants with multiple visualizations and asking them to select one based on a prompt. \rev{Unlike the previous two methods, which are open-ended, ``select a graph'' has limited exploratory power because its questions investigate the relative strength of a single affordance across a closed set of graphs}. For example, Fygenson et al. had participants report which of four visualizations makes a given message ``the most obvious" \cite{fygenson2023affordances}, and Xiong et al. and Levy et al. requested participants select \rev{the ``best''} visualization for different communicative goals from a set of options \cite{levy-tversky-gratuitous-graphics, xiong-afford-comparison}. 
 \rev{Select-a-graph methods facilitate larger sample sizes and easier reproduction than open-ended methods because its results can be quantitatively evaluated, which scales with less cost than qualitative review.}
\newline

\subsubsection{\textbf{Rating Construct Strength}}
Rating scales (e.g., Likert, visual analog, etc.) have been used to evaluate how strongly visual objects or treatments convey specific characteristics. For example, Ziemkiewicz and Kosara asked participants to rank visualizations of company departments on a scale from 1 to 5 depending on how likely the visualized companies were to embody different characteristics (e.g., stable, balanced, complete, rigid, etc.)\cite{Ziemkiewicz-dynamics-2010}. MacEachren et al. similarly requested participants rank visual encodings' (e.g., blur, grain, size) conveyance of uncertainty on a scale from 1 to 7 \cite{maceachren-semiotics}. Data measurement scales are a well-established tool and have also been used to elicit perceived difficulty and affective evaluations of visualizations, such as reported frustration or general workload (e.g., NASA's Task Load Index~\cite{nasatlx}). For a review of Likert scales in visualization evaluation, see \cite{south-likert-scale}.

The four methods we describe cover the entirety of \rev{cognitive} affordance-related investigations in visualization to the best of our knowledge. Given the variety of terminologies (\autoref{tab:terms}) used to describe the constructs studied by these methods, it is striking that only a small number of methods are commonly employed. This suggests significant overlap in the underlying cognitive processes associated with these constructs. We argue that \rev{cognitive} affordances can \rev{support} a unifying framework for more explicitly measuring, evaluating, and ranking the constructs described in this section.

\section{A Cognitive Affordance Framework for Visualization}
\label{sec:theory}
Section \ref{sec:existing_research} highlights \rev{visualization} research that theoretically or experimentally relates to \rev{cognitive} affordances, \rev{the results of} which we used as the empirical foundation of our proposed framework.
As shown in Figure \ref{fig:teaser}, our framework of cognitive affordances \rev{in visualization} includes three primary components: design decisions (Sec. \ref{sec:ind-vars}), readers (Sec. \ref{sec:readers}), and hierarchy of afforded information (Sec. \ref{sec:affordance-right}).

\subsection{Design Decisions}
\label{sec:ind-vars}
\rev{Cognitive} affordances \rev{in visualization} are driven, in part, by \textit{design decisions}, including but not limited to encoding choice, channels, arrangement of visual objects, labeling, annotations, and data shape~\cite{munzner2015visualization, Gitelman2019-Raw-Data} (Figure \ref{fig:teaser}, left). These elements coalesce to describe the entirety of a chart's visual output, and each has been shown to alter graph comprehension~\cite{Shah2002_graph_comprehension_book}. %
All design decisions could alter visualizations' \rev{cognitive} affordances, so compiling a list, as we do below, enumerates factors that may be influencing communicated information. 
 \begin{wrapfigure}{l}{0.2\textwidth}
\centering \includegraphics[alt={Top: two bar charts, each groups four bars into two groups of two bars. The left bar chart groups bars by position, leaving extra space between the groups, and the right bar chart groups by color, changing the color of the bars in different groups. Bottom: A bar chart of four values and a line chart of the same four values.}, width=0.20\textwidth]{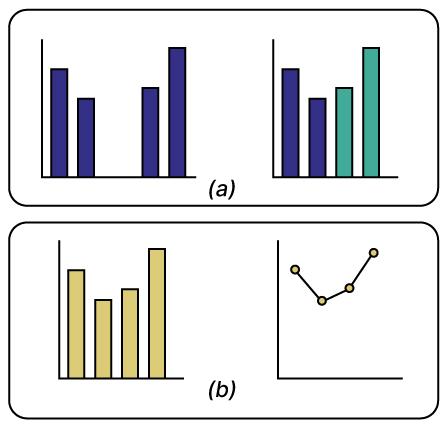}
    \vspace{-4mm}
    \caption{Examples of visual encoding manipulations. In (a) bars are grouped by position on the left, and color on the right. In (b) data is visualized with bar height and then point and line placement.}
    \label{fig:vis-encoding-ex}
    \vspace{-2mm}
\end{wrapfigure}
\textbf{Visual encoding} refers to the design choices that determine how underlying data is represented through visual elements, such as marks and channels. A comprehensive set of possible encodings can be informed by established taxonomies of visualization channels (e.g., \cite{munzner2015visualization}).Previous research has demonstrated that modifications to visual encoding significantly influence the information viewers extract from a visualization. 
For instance, Zacks and Tversky examined the differing information conveyed by lines versus bars \cite{zacks-tversky-bars-lines}, and Fygenson et al. explored grouping effects using position and color \cite{fygenson2023affordances} (see Figure \ref{fig:vis-encoding-ex}). Additional studies have investigated the intuitive associations between colors and concepts of ``more" or ``less" \cite{schloss-color-2019, golbiowska-color-intuitiveness}. 

\textbf{Visual arrangements}, the second design decision in Figure \ref{fig:teaser}, refer to the design of visual objects that do not affect their underlying data encoding strategy. \begin{wrapfigure}{r}{0.19\textwidth}
    \centering
    \includegraphics[alt={Top: two pie charts with the same slices in different order. Bottom: two pie charts with the same slices and ordering but rotated slightly differently.}, width=0.19\textwidth]{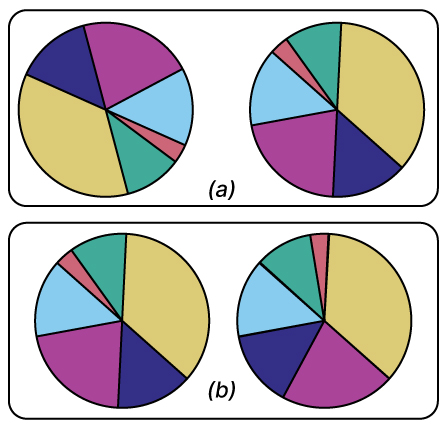}
    \vspace{-5mm}
    \caption{Examples of visual arrangement manipulations. (a) shows pie charts that are identical but slightly rotated. (b), shows pie charts with differently arranged slices.}
    \label{fig:vis-arrangement-ex}
    \vspace{-1mm}
\end{wrapfigure} For example, in Figure \ref{fig:vis-arrangement-ex}, the order in which categorical slices are positioned and the rotation of the pie chart do not affect the method of encoding. Visual arrangements can also include formatting decisions, such as changing the aspect ratio or resolution of a visualization.
Visual arrangements can be manipulated both within and between channels. For example, Ziemkiewicz and Kosara~\cite{Ziemkiewicz-dynamics-2010} explore how border thickness and color in part-to-whole visualizations affect perceived data dynamics, demonstrating \textit{between-channel arrangements}. \textit{Within-channel arrangements} include studies on bar positions in bar charts~\cite{xiong-afford-comparison, fygenson2023affordances} and the placement of filled icons in icon arrays~\cite{xiong-perceptual-biases-icon-arrays}. Similar to \textit{visual encodings}, a corpus of visual arrangements could be compiled from taxonomies of visual channels, such as those found in textbooks~\cite{munzner2015visualization, ware-info-vis}.

\begin{wrapfigure}{l}{0.25\textwidth}
    \centering
    \includegraphics[alt={top: two scatterplots with different titles. The right scatterplot reads ``Country X has the highest GDP'' and the left scatterplot reads ``Country X WINS GDP!''. Bottom: two scatteplors with differently shaped annotations pointing out the same outlier. The left scatterplot circles the point with a spiky shape, and the right scatterplot uses an arrow to point out the outlier.}, width=0.25\textwidth]{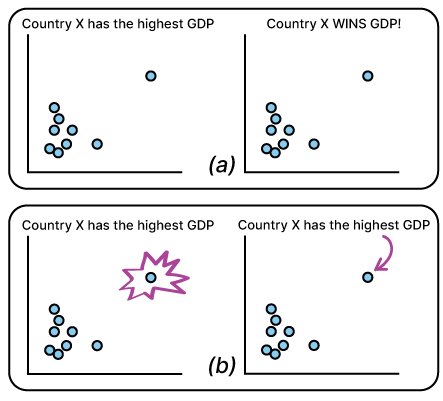}
    \vspace{-4mm}
    \caption{Examples of contextualizing visual elements: \rev{(a) title wording changes, (b) annotation shapes changes.}}
    \label{fig:cve-ex}
    \vspace{-2mm}
\end{wrapfigure}

\textbf{Contextualizing visual elements} (CVEs), the third type of design decision, are elements in a visualization that do \textit{not} encode data directly but provide useful information for interpreting visualized data. CVEs include titles, axis labeling, captions, embellishments, and annotations (see Figure \ref{fig:cve-ex}). %
Although not defined by a cohesive framework, evidence shows CVEs significantly influence how readers understand visualizations. For example, Stokes et al. found that adding annotations to visualizations impacts reader predictions of future trends \cite{stokes-annotations-bias}. 
Andry et al. explored how visual embellishments impact perceived salience and the order in which readers view different parts of visualizations \cite{andry-embellishment}. Kim et al. studied how caption content and visual saliency of chart features impact reader takeaways of line charts \cite{kim-charts-captions}. %
Other design decisions associated with CVEs include changing the position, color, density, and semantic content of CVEs\cite{stokes-annotations-bias, Bryan-eyetracking-vis-embellishments, andry-embellishment}.

\begin{wrapfigure}{r}{0.23\textwidth}
    \centering
    \includegraphics[alt={Top: a scatter plot on the left is approximated with a chart on the right that shows a mean line and shaded-in confidence interval. Bottom: The left figure shows a group of people icons that are all colored differently in one region of the icons. The right figure shows a group of people icons that are periodically alternating different colors.}, width=0.23\textwidth]{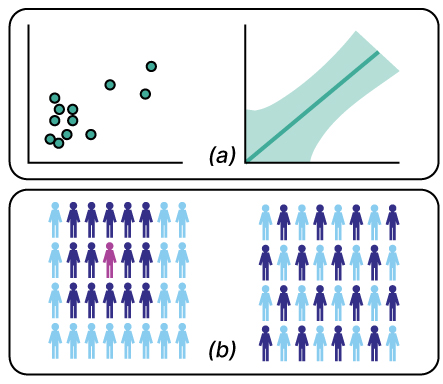}
    \vspace{-4mm}
    \caption{Examples of data manipulations: (a) individual dots in a scatter plot can be replaced with an aggregate mean and 95\% CI, (b) convenience vs systematic sampling.}
    \label{fig:vis-manipulations-ex}
    \vspace{-1mm}
\end{wrapfigure}

\textbf{Data}, the fourth type of design decision, encompasses choices about data selection, collection, aggregation, analysis, and context (see Figure \ref{fig:vis-manipulations-ex}).
Past work has shown that large outliers in data often are more salient than other visualized data points \cite{wolfe-vis-attention}, and that changing time series' y-axes from incident to cumulative occurrences can alter readers' takeaways \cite{padilla-covid-risk}. There is also evidence to suggest that showing averaged forecasts vs. individual conflicting forecasts impacts readers' affective response~\cite{padilla-mfvs-trust}. 
Data design decisions can include population sampling techniques (e.g., convenience, simple random, snowball, and cluster sampling \cite{FPH-sampling}). Exclusion criteria and analysis decisions (e.g., parametric assumptions, and Bayesian priors) are also data manipulations. Visual aggregation can also be a data manipulation (e.g., adding data averages, confidence intervals, or error bars).  
Although much related work in visualization investigates visual elements (e.g., encoding, arrangements, and CVEs), other visualization research has found substantial evidence that data shape can alter affective outcomes, such as perceived risk \cite{padilla-covid-risk} and trust \cite{padilla-mfvs-trust}, and performance variables, such as estimated averages \cite{moritz-avg-estimate-line}, perceived causation~\cite{xiong-illusion-causality}, and prediction accuracy \cite{padilla-mfvs-trust}. Although data shape may appear to be un-editable, common manipulations, such as removing outliers and downsampling or mathematically transforming data, can drastically impact shape.

\textbf{Situation}, which is the background and ancillary information surrounding a visualization, is the final type of design decision. Situation can alter visualizations' \rev{cognitive} affordances by directing and modulating readers' attention and guiding readers' attitudes. Situation design decisions have been studied by altering visualization readers' priming on specific topics, which can change the likelihood of them completing visual tasks~\cite{xiong-curse-knowledge}. Other situation decisions that could reasonably impact \rev{cognitive} affordances include changing the order of presented information in a narrative visualization, the framing of text accompanying a visualization, and the publisher or source of the visualization and its data. For example, in their survey of the perceived usefulness of visualizations in rural Pennsylvania, USA, Peck et al. found that individuals' ranking of visualization usefulness was impacted by their attitudes about the visualization publisher~\cite{Peck-data-personal-rural}.

Design decisions change visualizations' communication of information, primarily by altering visual elements, and are a driving factor of which, and to what degree, information is afforded, and thus are foundational to the discussion, study, and utilization of visualization's \rev{cognitive} affordances. We place design decisions on the far left-hand side of Figure \ref{fig:teaser}, as they drive the creation of \rev{cognitive} affordances, which are susceptible to moderation based on reader characteristics, and are detailed in the next section.

\subsection{Readers}
\label{sec:readers}
\textit{Reader characteristics} act as moderating variables (Figure \ref{fig:teaser}, middle), potentially enhancing, suppressing, or removing \rev{cognitive} affordances. %
For example, visualization comprehension and performance can vary based on readers' background, cognitive function, and prior beliefs~\cite{Ziemkiewicz-individual-users, liu-individual, xiong-illusion-causality, xiong-curse-knowledge}. When evaluating, theorizing, and designing for \rev{cognitive} affordances, it is important to document these characteristics, although some may have a greater impact on affordances than others. For instance, factors such as exposure to data and low vision are likely to influence visualization interpretation more significantly than gender. We outline four categories and provide various examples of each to offer a comprehensive view of variables that may moderate \rev{cognitive} affordances. We do not suggest that effective research requires collecting all possible characteristics, but aim to highlight those that past research indicates are more likely to impact interpretation.

\textbf{Demographics}, including age, gender, and language use, are established in cognitive psychology to potentially impact individuals' perceptions. Within visualization, there is also evidence that these variables can correlate with readers' takeaways. For example, Castro-Alonso et al. found that gender can correlate with readers' ability to learn from dynamic visualizations \cite{castro-alonso-gender}, and Ziemkiewicz and Kosara found that preference for visual metaphors, and differing approaches to sensemaking in visualization, can vary with self-reported gender \cite{Ziemkiewicz-individual-metaphor}. Tversky et al. also found that the native language of children impacts the direction they attribute to naturally expressing the passage of time\cite{tversky-time-series}), and Panavas et al. showed that children interpret visualizations less accurately than adults~\cite{panavas-children}.

\textbf{Learned skills} refer to readers' abilities that may aid or sway their understanding of information, visualization, numbers, or data. Importantly, learned skills are not implicit traits but rather abilities that can be augmented with additional training. Education level, occupation, expertise, or graphic/numeric literacy are not predetermined or static attributes, and past work has shown their potential to vary with visualization preferences and interpretations \cite{vazquez-ingelmo-expertise, shah-1995-comprehending-line-graphs, shah-bar-line-graph-comprehension, shah-graph-comprehension-prior-knowledge, culbertson-1959-graph-comprehension}. 

\textbf{Prior beliefs}, the third reader characteristic in Figure \ref{fig:teaser}, includes political alignment, religious affiliation, institutional trust, and preconceived metaphors, and can contribute to a visualization\rev{'s cognitive} affordances. For example, Luo and Zhao showed how political alignment affects readers' attention and perception of identical climate change evidence \cite{luo-attention-climate-change}, and Yang et al. found that encoding choice when visualizing Republican candidates' polling can have an impact on participants' intentions to vote \cite{yang-election-forecast}. Peck et al. interviewed readers in rural Pennsylvania, U.S.A., and found that political affiliation and preconceived notions about publishing sources and topics can impact readers' ranking of visualizations' usefulness \cite{Peck-data-personal-rural}. Although intentions to vote and perceived usefulness are not \rev{cognitive} affordances, per se, these works suggest that interpreted information can vary with prior beliefs. Similarly, Xiong et al. demonstrated that preconceived biases can impact visualization comparisons and perception of correlation and causality \cite{xiong-illusion-causality,  xiong-belief-biases-correlation}. Lastly, Ziemkiewicz and Kosara found that aligning visualization methods with readers' preconceived visual metaphors can impact readers' accuracy and preferences when interpreting visualizations \cite{Ziemkiewicz-individual-metaphor}.

\textbf{Physical and cognitive function}, the final reader characteristic, refers to inherent abilities such as color vision deficiency, working memory, visual-spatial ability, and locus of control, all of which affect how readers understand and make sense of visualizations.
In their research on spatial abilities' effect on visualization interpretation, Chen and Czerwinski found that readers with ``high" spatial ability tended to employ hierarchical task strategies, whereas readers with ``low" spatial ability were more likely to use brute force methods \cite{chen1997spatial}. Chen and Czerwinski's work serves as a precedent that spatial ability may impact cognitive affordances. Ziemkiewicz et al. found evidence that individual differences in locus of control (i.e., ``a person's tendency to see themselves as controlled by or in control of external events" \cite{Ziemkiewicz-locus-of-control}) impact individuals' speed, accuracy, and preference when completing tasks with visualizations. Although Ziemkiewicz et al. investigated only performance and preference, their work motivates future research on locus of control and \rev{cognitive} affordances. Lastly, \textit{working memory}, a cognitive process that enables individuals to retain a limited amount of information for a short period, has also been linked to differences in visualization interpretation~\cite{adams2018theories}. Working memory is responsible for engaging and controlling attention~(for extended definitions, see \cite{cowan2017wm, padilla-decision-making}), and prior work has found that individuals with high working memory capacity have greater improvements in decision-making tasks when using quantile dot plots than other readers~\cite{castro2021_workingmemory}.

Individual reader characteristics can moderate the \rev{cognitive} affordances that design decisions create. Although it may be impractical for visualization designers to plan their designs to appeal to all audiences, the knowledge that afforded information can change across visualization readers can help designers create with more intention. Acknowledging reader characteristics' role in visualizations' affordances can also support more inclusive design. For example, visualization best practices advocate for color-vision-deficit-friendly palettes and/or double encoding to support the affordance of information to readers with varying levels of color vision\cite{oliveira-cvd-2013, white_2017_cvd}.

\textbf{Collecting reader characteristics} can be supported through questionnaires (e.g., demographic, working memory\cite{vallat-azouvi-workingmemory, workingmemory-conway}, locus of control\cite{rotter-locus-1966, ferguson1993rotter-locus}, and graph literacy \cite{okan-sgliteracy}), as is common practice in HCI~\cite{muller2014surveyhci}. We focus our methodology discussion on techniques that are found in \rev{cognitive} affordance-related work in visualization. HCI papers on survey research \cite{muller2014surveyhci} can provide further guidance. 

\textbf{Interaction of reader characteristics} is also possible, resulting in many combinations of factors that could impact afforded information. Additionally, a reader's demographic information, such as salary, education, or race can correlate with their learned skills and prior beliefs. Our understanding of visualization\rev{s' cognitive} affordances would be greatly supported by work that investigates the correlations and interaction of these characteristics.

\subsection{Afforded Information}
\label{sec:affordance-right}
The cognitive actions that are afforded by design decisions and reader characteristics manifest as \textit{afforded information} on the right side of our framework (Fig. \ref{fig:teaser}). Afforded information is a broad description of any form of takeaway or understanding about data that a visualization communicates, including visual tasks (e.g., compare bar A to bar B), messages (e.g., category A is larger than category B), characteristics (e.g., the data shown is relatively uncertain), and more. All of these forms of information can be found when reviewing the measured outcomes of the work discussed in Section~\ref{sec:existing_research}. We abstain from delineating different types of afforded information and refer interested readers to current literature on visualization tasks, messages, and takeaways, including Brehmer and Munzner's multilevel visualization task typology \cite{brehmer-munzner-task-taxonomy}, Quadri and Rosen's extensive review of perception-based visualization tasks \cite{quadri-perception-survey}, Amar et al.'s discussion of low-level components of activity in visualization \cite{amar-task-taxonomy}, and Friel et al.'s summarization of the types of questions that can be asked during graph comprehension \cite{friel-graph-sense}.

Visualizations' afforded information stems from design decisions that influence the likelihood of readers performing visual or cognitive tasks. Although every visualization typically affords a multitude of information, we posit that design decisions and reader characteristics can make some pieces of information more probably communicated than others. For example, although two readers may glean different pieces of information from a chart that uses blurry marks, past research shows evidence that both are likely to find that the blurred marks communicate uncertainty \cite{maceachren-semiotics}. The congruence of afforded information and intended information may impact whether a designer feels a visualization is ``successful'' \cite{quadri-doyouseewhatisee}. Our framework can scaffold reasoning about likely interpretation, and structure experiments to investigate the impact of visualization design decisions.
           
\subsubsection{\textbf{Hierarchy of Affordances}}
\label{sec:hierarchy}

 Both visualization and non-visualization, theoretical and experimental works provide evidence that within a total set of afforded \rev{tasks or} information, affordances can exist to varying degrees \cite{davis-chouinard, mcgrenere-ho-aff-2000, franchak-adolph-aff-2014, xiong-curse-knowledge, fygenson2023affordances}. We propose two tenets that describe this fluid hierarchy:

\textit{\textbf{T1) A visualization affords a finite amount of information to its reader.}}
This set is equivalent to the traditional definition of affordance by Norman \cite{norman-1999, Norman_revised} (i.e., information either is or is not afforded, depending on whether it is perceived as possible to extract). This total set of afforded information is visualized in green on the right side of Figure \ref{fig:teaser}.\begin{wrapfigure}{r}{0.23\textwidth}
    \centering
        \vspace{-4mm}
\includegraphics[alt={Two bar charts that groups four bars into two groups of two. The left chart has bars in each group colored differently. The right chart has bars in a single group colored the same.}, width=0.23\textwidth]{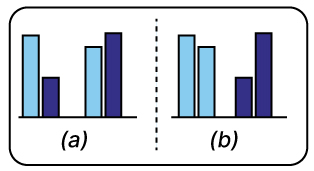}
    \vspace{-6mm}
    \caption{\rev{(a) When bars from different color groups are placed near each other, across-group comparisons are more obvious~\cite{fygenson2023affordances}. (b) When bars within color groups are placed together, within-group comparisons are more obvious~\cite{fygenson2023affordances}.}}
        \vspace{-2mm}
    \label{fig:bar-spacing}
\end{wrapfigure} 
In contrast, information that is not afforded is written in gray below the green text. As design decisions change, or as individual reader characteristics moderate the effects of these design decisions, previously not afforded information may become afforded, and vice versa.

\textit{\textbf{T2) Some information is more strongly afforded than others.}}
Affordances are relative, as shown on the right side of Figure~\ref{fig:teaser}, where more strongly afforded information is higher up and in a darker green. 
For example, Figure~\ref{fig:bar-spacing} shows two bar charts that encode the same data with different bar placement. This placement can change readers' perception of intended communication, which would rearrange the vertical order, i.e., the hierarchy, of \rev{cognitive} affordances in Figure~\ref{fig:teaser}.

\begin{figure}[h!]
    \centering
    \includegraphics[alt={A flow approximating Figure 2, where a grouped bar chart and a set of line charts occupy the Design Decisions box, the text ``undergrad students'' occupy the readers oval, and two different pieces of information switch vertical ordering in the Hierarchy of Afforded Information box. The grouped bar chart has an arrow that leads to ``compare: bars within year'' higher than ``compare: compare trend across years''. The line charts have an arrow that leads to ``compare: compare trend across years'' higher than ``compare: bars within year''.}, width=0.50\textwidth]{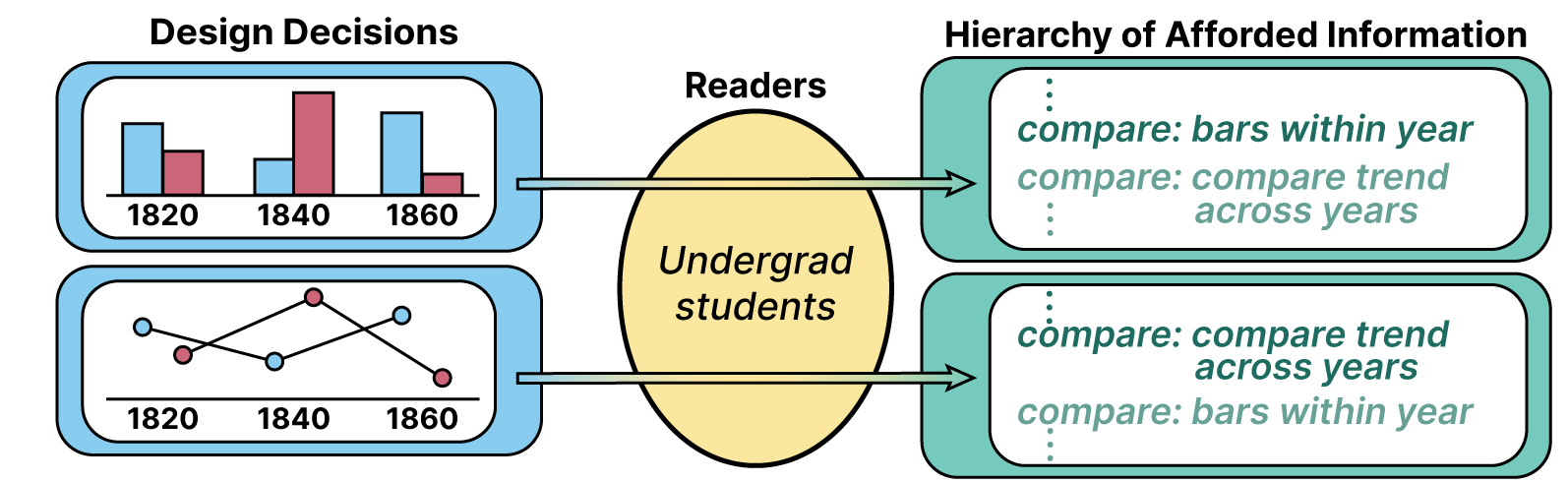}
    \vspace{-5mm}
    \caption{A result from Shah et al. \cite{shah-graph-comprehension-1999} visualized in our framework. \rev{Undergrad students are more likely to compare data within a year instead of data over time when shown grouped bar charts. The opposite is true when students are shown the same data via line charts.} This result is an example of how design decisions can impact a visualization's hierarchy of \rev{cognitive} affordances.}
    \label{fig:Shah-results}
    \vspace{-1mm}
\end{figure}

This relational ranking of \rev{cognitive} affordances can be seen in past visualization work. For example, many of the visualization experiments reviewed in Section~\ref{sec:rel-constr} do not generate singular results from all participants, but rather find a spectrum of results with varying strengths or signals~\cite{xiong-afford-comparison, fygenson2023affordances, zacks-tversky-bars-lines, shah-graph-comprehension-1999}. This ranking is often not made explicit in the results, discussion or conclusion sections of papers. Rather, authors tend to focus their experimental conclusions on the strongest results, reporting only the most prominent \rev{cognitive} affordances, and potentially perpetuating a false binary. In turn, experimental findings that indicate the evidence of less strong affordances can be easily overlooked. In Figure~\ref{fig:Shah-results}, we illustrate results from Shah et al.~\cite{shah-graph-comprehension-1999} \rev{to show how our framework can structure results to more fully report findings.}

\begin{figure}[h!]
    \centering
    \includegraphics[alt={A flow approximating Figure 2. A grouped bar chart that labels each group of bars A1 and B1, A2 and B2, and A3 and B3 occupies the Design Decisions box. Two different reader ovals show the text ``data vis experts'' and ``non-experts'', and two different pieces of information switch vertical ordering in the Hierarchy of Afforded Information box. The data vis experts oval has an arrow that leads to ``compare: A1 vs B1'' higher than ``compare: average of As vs average of Bs''. The non-experts oval has an arrow that leads to ``compare: average of As vs average of Bs'' higher than ``compare: A1 vs B1''.}, width=0.50\textwidth]{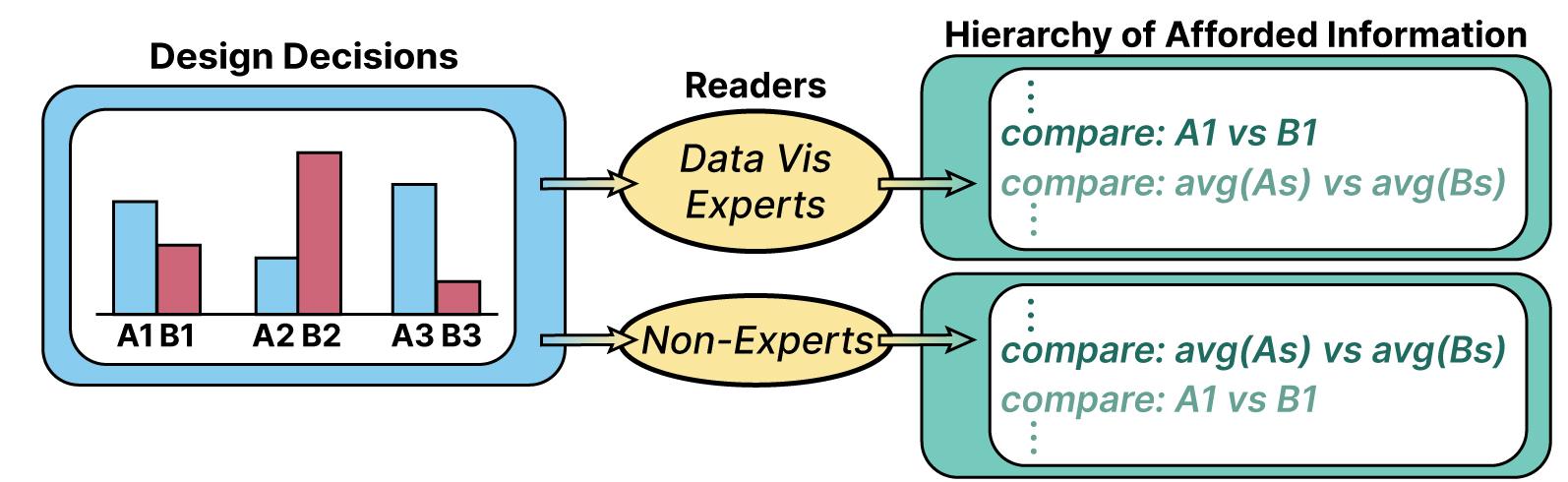}
    \vspace{-5mm}
    \caption{A result from Xiong et al. \cite{xiong-afford-comparison} visualized in our framework\rev{. Experts and nonexperts have different rankings of afforded information when shown the same bar chart.} This result shows how reader characteristics can impact a visualizations' hierarchy of \rev{cognitive} affordances.}
    \label{fig:xiong-results}
    \vspace{-3mm}
\end{figure}

We can see both tenets in action in Xiong et al.'s experimental results \cite{xiong-afford-comparison}. Xiong et al. tested how strongly messages of comparison are afforded given different bar \rev{placement} in bar charts and participant populations with different levels of data visualization expertise. In Figure~\ref{fig:xiong-results}, we use our framework to show that Xiong et al. found that manipulating spatial arrangements \rev{(a design decision)} of bars results in different \rev{cognitive} affordances depending on readers' experience with data visualizations (a Learned Skill characteristic). 
\vspace{-3mm}

\section{Practical Application of the Framework}
\label{sec:framework-application}
Beyond outlining theoretical foundations, our framework provides a structure for practitioners to evaluate visualization drafts, and serve as scaffolding for the compilation of \rev{cognitive} affordance findings \rev{in visualization}. In the following sections, we exhibit how our framework can be used for both scenarios.

\subsection{Comparison of Design Decisions}
\label{sec:comparing-design-decisions}
The \rev{cognitive} affordance framework can help designeres refine their visualizations. %
Consider a designer at an electronic health record company, such as MyChart\footnote{\url{https://www.mychart.org/}}, who is creating a visualization to help patients understand trends in their self-reported symptoms over time. These self-reported variables, known as patient-reported outcomes (PROs), have proven beneficial for clinical care and patient outcomes~\cite{Crins_2020, Deyo_2015, penedo-2020}.
\begin{figure}
    \centering
    \includegraphics[alt={A set of steps titled ``Redesigning Visualizations with Cognitive Affordances''. The first step reads ``What do I want to prioritize communicating?'' and shows the framework from Figure 2 but with only the Afforded Information box filled in with ``Health outcomes over time'' and ``Symptoms improving or worsening''. The second step reads ``To whom do i want to prioritize communicating?'' and shows the same figure as in step 1 but fills in the Readers oval with ``Patients using EHRs'' with traits ``many ages'' and ``many levels of graph literacy''. Step 3 reads ``What does my current design afford?'' and shows a line chart going up, with a y-axis labeled ``more pain'' on the top and ``less pain'' on the bottom, and an x-axis labeled ``time.'' Below the line chart is the same figure as in step 2 but replaces the afforded information piece that used to read ``Symptoms improving or worsening'' with ''positional semantics or visual position convention''. This piece of information is marked with a large red X to indicate it is not an ideal affordance. Step 4 reads ``what is making the current design afford undesired info?'' and shows the same framework figure as in step 3 but fills in the design decisions box with two encodings labeled ``time = position along x-axis'' and ''positional encoding along y-axis''. The first encoding leads to the afforded information of ``health outcomes over time'', whereas the second encoding leads to the unideal information of positional semantics or visual position convention. Step 5 reads ``what design decisions afford what we want?'' and shows color, pictograph, and shape as encodings in the design decisions box that lead to different afforded information. Of these three encodings, only color and pictograph lead to ideal afforded information. Step 6 reads ``redesign using these design decisions'' and shows a line chart with smiley face pictographs as its dots, a line chart with red-yellow-green colored dots, a line chart with red-yellow-green smiley faces as its dots, and timeline with red-yellow-green smiley faces.}, width=.97\linewidth]{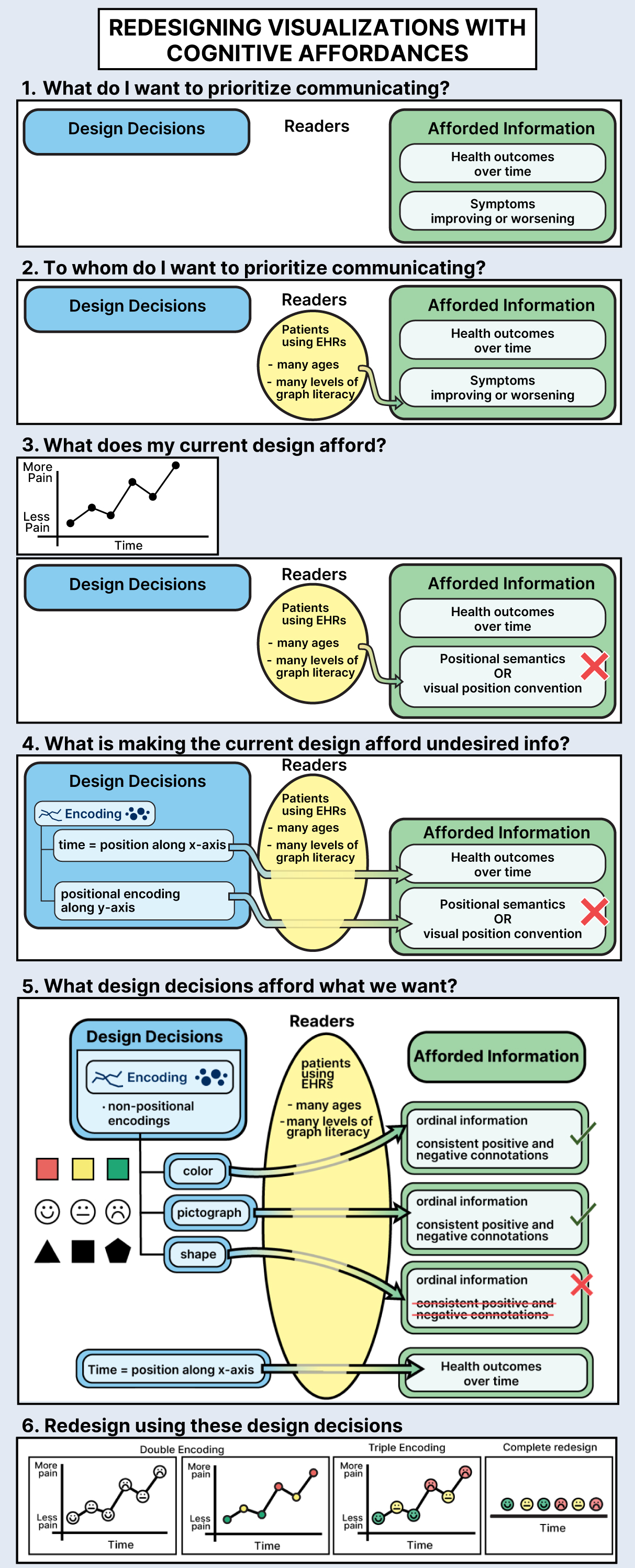}
    \vspace{-3mm}
    \caption{Workflow for examining and re-designing a visualization using cognitive affordances. \rev{In steps 1 and 2, a designer populates the cognitive affordance framework with their communication goals and target audience. In step 3, the designer uses investigative methods or past research to determine what their current visualization affords. In step 4, the designer identifies undesired information that their visualization communicates and uses cognitive affordances to trace the information back to root design decisions. In step 5, the designer uses cognitive affordances to hypothesize about the likely communication of alternate design decisions. In step 6, the designer implements visualization redesigns using step 5's identified encodings.}}
    \label{fig:ex-flowchart}
\end{figure}
\indent\textbf{Heuristic evaluation with current methods}: First, we will examine the designer's visualization process using current frameworks and conventions. The designer's visualization goal is to communicate a variable over time. Following convention, the designer decides to show a PRO time series with a line chart. 
Line charts are typically used to communicate longitudinal data \cite{beniger-stats-history-1978} and are a common visualization method for PROs \cite{Brundage2005, Brundage2015, Brundage2018, McNair2010, Smith2016, Tolbert2018, Snyder2017, Snyder2018}. This decision aligns with the hierarchy of expressiveness and effectiveness \cite{munzner2015visualization}, which ranks position encodings highly. The designer first chooses to orient the y-axes so they are semantically congruent, adhering to the ``more is up" principle \cite{lakoff-emboided-cog} (see line chart in Fig. \ref{fig:ex-flowchart}, step 3). After reviewing the graph with a colleague, the designer discovers that their colleague is confused by this y-axis orientation and recommends aligning the y-axes with a common positional convention in which higher positions correspond to more ideal or ``happier" outcomes \cite{lakoff-emboided-cog}. If the designer wants to consult visualization literature, they may come across a few resources (e.g., \cite{padilla-woodin-metaphor-graphical-convention}), with mixed results when evaluating y-axis directionality from a performance perspective. The designer might consult cognitive fit or graph comprehension frameworks, which emphasize aligning visualizations with readers’ mental schemas, and can explain why a design lacks alignment, but offer little actionable guidance on how to improve it.

\textbf{Heuristic evaluation with the Cognitive Visualization Affordance Framework}: If the designer, however, were to turn to \rev{cognitive} affordances, they would find not only resources to outline the \textit{root cause} behind confusion, but also a \textit{logical framework} for addressing these causes and creating a better visualization. \rev{After identifying their communication goals, the designer could use cognitive affordances to better understand how their line chart misaligns with those goals, and what other design decisions may make reaching those goals more likely. In more detail,} to use the framework to support their design process, first the designer would enumerate the information they want to communicate with their visualization (Figure \ref{fig:ex-flowchart}, step 1). Next, the designer would specify reader characteristics that are likely to be present in their target audience (Figure \ref{fig:ex-flowchart}, step 2). This specification can be supported by classic methods in user experience design, such as creating user personas \cite{Pruitt2003Personas, sundt2017user}. The designer would then examine their visualization to identify undesired affordances and any missing desired affordances (Figure \ref{fig:ex-flowchart}, step 3). Currently, \rev{cognitive} affordances of PRO line charts are most informed by qualitative experiments published in informatics journals (e.g., \cite{Brundage2018, Brundage2015, Snyder2018, Snyder2017, Tolbert2018}). These experiments adhere to the describe-a-graph approach discussed in \ref{sec:methods}, and many are conducted on participants, such as cancer patients, who are likely to share reader characteristics with the designer's audience. If the designer did not have resources on known \rev{cognitive} affordances of their design, they could conduct their own study using one or more of the methods in section \ref{sec:methods}.

Investigating PRO line charts' \rev{cognitive} affordances would lead the designer to recognize that the positional encoding of symptoms on the y-axis can afford a semantic mapping (``more is up") to some readers and a positionally conventional mapping (``better is up") to others (Figure \ref{fig:ex-flowchart}, step 4)\rev{, and thus} the root cause of confusion with their line chart is encoding health outcomes via y-axis position.

Now, the designer can make an informed redesign. 
Rather than simply flipping the y-axis based on a colleague's preference \rev{or brute-force testing other encodings}, the designer can recognize that a positional y-axis encoding confuses some \rev{readers} and can explore alternatives that better align with their desired affordances. For example, in step 5 of Figure \ref{fig:ex-flowchart}, the designer can enumerate and evaluate nonpositional encodings' affordances and determine \rev{that shape is not an appropriate alternate, whereas color and pictographs both} offer affordances in line with the designer's goals. Notably, a \rev{cognitive} affordance framework can suggest redesigns that are not supported from a precision-based lens, such as those shown in step 6 of Figure \ref{fig:ex-flowchart}, providing logical solutions that might otherwise be overlooked.

This walkthrough presents a theoretical solution to the well-documented problem of confusing PRO line charts \cite{Brundage2015, Snyder2018}. Further validation of our framework and its usefulness could investigate the effectiveness of affordance-motivated design solutions. Although this experimental validation is outside the scope of our paper, it presents an avenue through which future work may validate our framework.

 \subsection{Organization of Research Findings}
\begin{figure}[h!]
\centering  
    \includegraphics[alt={Rows separated into 3 columns: design decisions, reader characteristics, afforded information. The top row is labeled Zacks \& Tversky 1999 and in the Design Decisions column it shows Encodings: bars, lines and Data: discrete, continuous, in the Reader column it shows Learned skills: undergraduates, and in the afforded information column it shows ``lines afford trend assessments'' and ``bars afford discrete comparisons''. The bottom row is labeled Shah et al., 1999 and has the same info as the top row, except the Design Decisions encodings read Data: percent, absolute, and there is an additional piece of afforded information in the rightmost column that reads ``percent vs absolute scaling doesn't impact affordances.''}, width=0.48\textwidth]{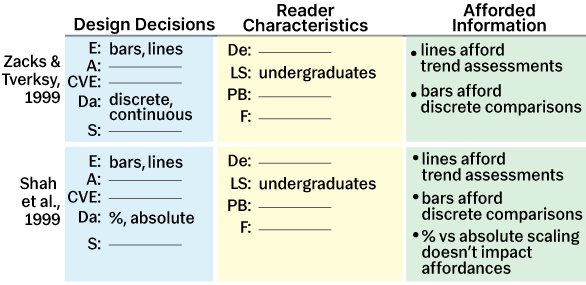}
    \vspace{-2mm}
    \caption{Our framework can support easier comparison of two \rev{cognitive} affordance findings, \rev{such as those from Zacks \& Tversky and those from Shah et al., shown above}. E = Encoding, A = Arrangement, CVE = Contextualizing visual elements, Da = Data, S = Situation, De = Demographics, LS = Learned skill, PB = Prior beliefs, F = Physical \& cognitive function.}
    \label{fig:table-comparison}
\end{figure}

The \rev{cognitive} affordance framework also provides structure for compiling visualization research findings for easier reference and exploration. Firstly, by outlining the impacting factors of \rev{cognitive} affordances, this framework facilitates easier synthesis. For example, past work, such as Zacks and Tversky's study of bars and lines~\cite{zacks-tversky-bars-lines}, could be categorized for faster juxtaposition against similar work, such as Shah et al.'s investigation of bars and lines~\cite{shah-graph-comprehension-1999}, as seen in Figure \ref{fig:table-comparison}.

\begin{figure}[h]
\centering  
    \vspace{-1mm}\includegraphics[alt={Top: Query box that reads `` Return visualization affordance findings where: design decisions have encodings = line and data = PRO scores over time and where reader characterstics have ages that includes people between 50 and 65 years old and with education levels of less than graduate school. Bottom: Output box with four papers that match the query parameteres and their tested design decisions and reader characteristics and discovered afforded information.}, width=0.5\textwidth]{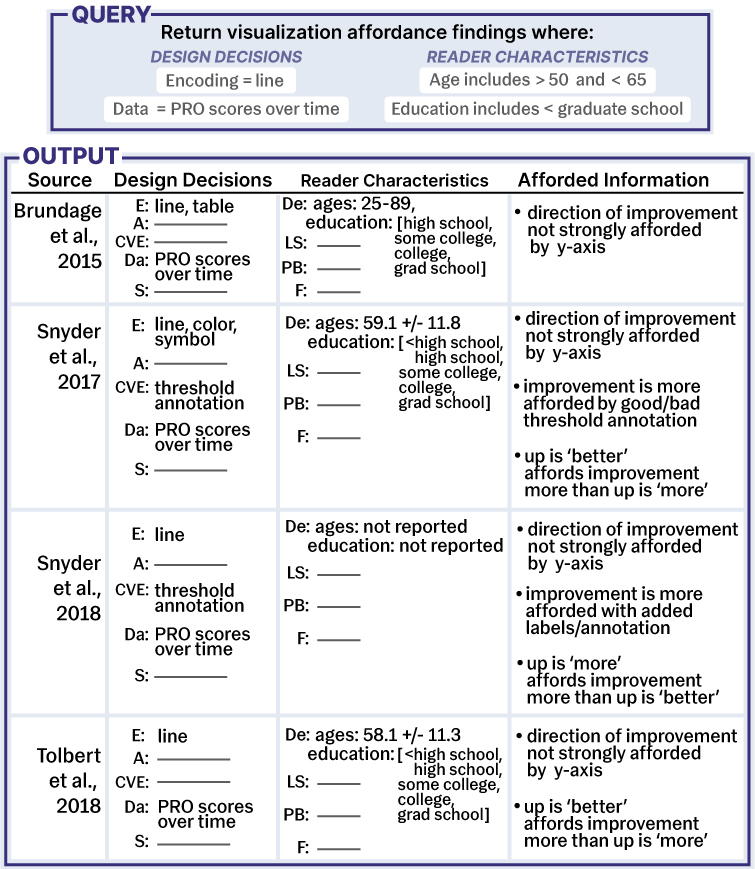}
    \vspace{-5mm}
    \caption{Example of a query and resulting output from a database \rev{structured} on our \rev{cognitive affordance framework. A user may want to survey research on the cognitive affordances of encoding patient reported (PRO) scores with lines for middle-aged readers who have not gone to graduate school.}}
    \label{fig:query-table}
    \vspace{-3mm}
\end{figure}

A quick comparison of these experiments' tables allows us to see similarities in tested encodings (E: bars, lines), differences in data manipulations (Da), similarities in tested readers (LS: undergraduates), and consistencies across found affordances (lines afford trends, bars afford discrete comparisons). 
These tables also provide a structure that could be extrapolated into a database, facilitating compilation of results and large-scale querying, which may prove useful to practitioners who need to investigate specific design decisions and possible alternatives. For example, if the designer from  Section \ref{sec:comparing-design-decisions} wanted to compare research on PRO line charts' affordances, our framework could support a queryable database that returns a table of \rev{cognitive} affordances, as shown in Figure \ref{fig:query-table}.

\vspace{-2mm}

\section{Discussion}
\label{sec:discussion}
We translate \rev{cognitive} affordance theory to a visualization context, presenting a framework to support more purposeful design. Below, we discuss how \rev{cognitive} affordances can inform understanding of unintentional communication, and relate to visualization evaluations of precision and ease.

\subsection{Unintentional \rev{Cognitive} Affordances}
\label{sec:unintentional} 
\begin{wrapfigure}{r}{0.23\textwidth}
\vspace{-3mm}
    \centering
\includegraphics[alt={A circular dot plot that is mostly navy blue with some lighter white-blue dots here and there. A legend at the bottom of the plot reads ``percent concentration of sea ice'' where 50 correlates to a light white-blue color and 100 corresponds to a dark navy.}, width=0.23\textwidth]{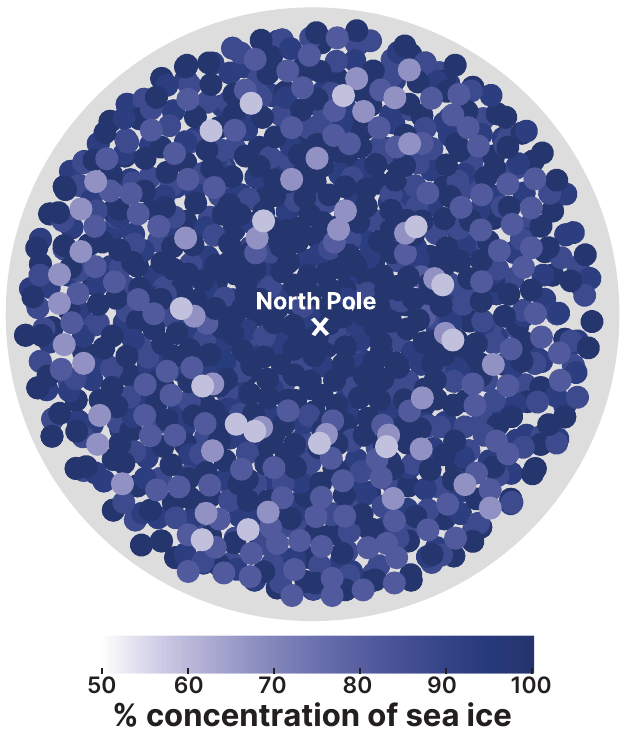}
\vspace{-3mm}
    \caption{Recreated from TowardsDataScience.com~\cite{sea-ice-tds}, this plot encodes \rev{higher} concentrations of sea ice \rev{in darker blue, despite ice appearing white in lived experience}.}
    \label{fig:ice_heatmap}
    \vspace{-6mm}
\end{wrapfigure}

\rev{Cognitive} affordances do not always align with designers' goals~\cite{quadri-doyouseewhatisee}. In fact, it is possible for visualization design decisions to mislead or confuse readers. One can expect interpretation problems when there is a mismatch between ideally communicated information and a visualizations' afforded information. A common example \rev{occurs when} color encodings do not semantically relate to data \cite{schloss-color-2019, lin-color-semantics-algo}. Take Figure~\ref{fig:ice_heatmap}, which shows a color-encoded dot map of the percent concentration of ice in the Arctic circle. Although English-speaking readers tend to find darker colors afford more of something (i.e., a dark-is-more bias) \cite{Silverman_Gramazio_Schloss_2016}, ice is also associated with a white/lighter color due to lived experience (e.g., a white iceberg in a blue sea). Using navy to encode areas of high ice concentration can afford much more seawater (and much less sea ice) than the data shows.
 Understanding visualization\rev{'s cognitive} affordances can help practitioners avoid creating designs that confuse or mislead.

\subsection{Precision, Ease, and \rev{Cognitive} Affordance}
\label{sec:ease}
The most popular methods of evaluating visualizations investigate how accurately or quickly readers perform tasks, often motivated by the theory that these metrics are proxies for cognitive effort\cite{quadri-perception-survey}. Optimizing visualization designs to reduce effort has furthered our field, creating hierarchies of channels \cite{munzner2015visualization} and facilitating the development of visualization recommendation systems \cite{vizql-hanrahan, polaris-stolte}. The prolificity of these methods begs the question: What do current methods lack, and why is \rev{cognitive} affordance necessary?

\indent One answer to this question is that the majority of precision and ease research investigates the impact of visualization design on the completion of specific tasks, whereas realistically readers may not approach visualizations with predetermined tasks, but rather in an exploratory fashion without strict intentions. In traditional precision and ease evaluation, tasks are treated as an input, but in practice, tasks are often the result, or output, of reader interactions with visualizations. \rev{Cognitive} affordances seek to uncover which tasks are initiated by readers when they are exposed to a visualization, rather than which tasks are best facilitated.\begin{wrapfigure}{l}{0.25\textwidth}
\vspace{-2mm}
    \centering
\includegraphics[alt={Two scatterplots with positively correlated data roughly along the line x = y and x and y axes that range from 1 to 4. The plot on the left has a dot on the point (0,1), and the title of the graphs reads ``Compare the Trends!''}, width=0.25\textwidth]{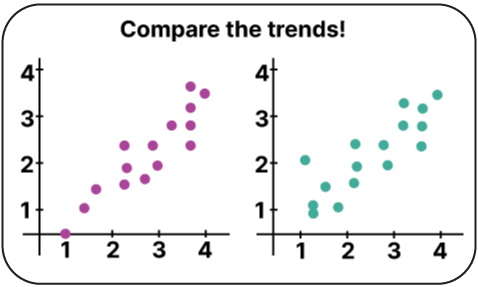}
    \vspace{-7mm}
    \caption{Although it is easy to read the value from the left graph's lowest data point, is that the first piece of information these charts communicate? \rev{The title encourages readers to focus on the graphs' trends.}}
    \label{fig:ease-aff}
    \vspace{-1mm}
\end{wrapfigure}
We must also recognize the distinction between the ease of extracting information and the affordance of information. Fundamentally, \rev{cognitive} affordance is indicative of implication and intention. Readers may be able to easily extract one piece of information from a visualization, while also recognizing that visualization's intent is to communicate different information. Take Figure~\ref{fig:ease-aff}, for example. If a researcher were to instruct a reader to extract the smallest x-y point value in the left-hand chart, a reader could quickly and accurately determine the point was at (1,0) \cite{cleveland-mcgill-1987}. At the same time, a reader who is presented with the charts and no other information would likely read the title (a Contextualizing Visual Element in our framework) and conclude the charts' intention is to imply a comparison of trends between scatterplots \cite{kong-title-slant}.

\begin{wrapfigure}{r}{0.2\textwidth}
 \vspace{-2mm}
    \centering
\includegraphics[alt={Two scatterplots with the same points, the plot on the left has a line that connects all of its dots.},width=0.2\textwidth]{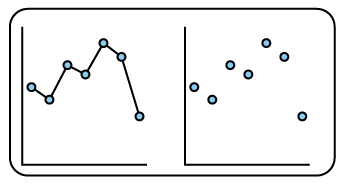}
    \vspace{-5.5mm}
    \caption{Connecting a scatterplot's points affords continuity and thus a single entity \cite{tversky-visualizing-thought}.}
    \label{fig:connected-ex}
    \vspace{-3mm}
\end{wrapfigure}

Additionally, \rev{cognitive} affordance investigates the communication of information \textit{beyond} raw data. For example, the charts in Figure~\ref{fig:connected-ex} likely require equal effort to communicate the relationship between graphed variables~\cite{cleveland-mcgill-1987}. However, the left chart affords a single entity changing over time, whereas the right chart affords many different entities~\cite{tversky-visualizing-thought}.  Without examining the extent of readers' perceived information, we cannot fully understand visualizations' communication. We design the previous examples to illustrate where performance metrics may not fully describe an audience's experience with a graph, and where \rev{cognitive} affordances can offer relevant insights. In practice, however, it is possible that performance and \rev{cognitive} affordance are strongly correlated. The potential for this correlation is only emphasized by the success with which performance metrics have guided visualization theory and practice thus far. To further investigate this correlation, we must strengthen the conceptualization and examination of \rev{cognitive} affordances in visualization.

\section{Future work and Limitations}
\label{sec:future-work-limitations}
We hope this work will act as a starting point for future \rev{cognitive} affordance research in \rev{visualization}. The framework we present is an initial scaffold to demonstrate the process by which visualization\rev{s' cognitive} affordances are created and altered. Future work \rev{could lead to developing our framework into a more comprehensive theory} and further validating its accuracy and usefulness through evaluations of affordance-based visualization designs, like those presented in Figure \ref{fig:ex-flowchart}. Additionally, although there is evidence in both affordance and visualization research to support reader characteristics' impact on visualization\rev{s' cognitive} affordances, further exploration as to how reader characteristics impact each other and specific affordances would add dimension to our framework.

This paper joins a growing body of research that advocates for evaluating visualizations beyond response time and accuracy \cite{lee-robbins-affective-vis, borkin-memorability}. We integrate past research to highlight the importance of \rev{cognitive} affordances \rev{in visualization} and provide a framework to support \rev{design reasoning} beyond precision-based metrics. Continuing to investigate \rev{cognitive} affordances in visualization will enhance our understanding of how visualizations communicate, advance visualization theory, and inform actionable design recommendations. The development of infrastructure, such as a catalog of \rev{cognitive} affordances \rev{in visualization} would support this investigation. %

\section{Conclusion}
\rev{Cognitive} affordances inform how design decisions impact the likelihood of communicating pieces of information to visualization readers. Research on this topic can warn practitioners when a visualization is likely to mislead its audience. %
We present a framework to translate cognitive affordances into a visualization context. In doing so, we \rev{contextualize cognitive affordances within} past research, explain practical applications of our framework, and discuss avenues of future research.

\section*{Acknowledgments}
We would like to acknowledge our incredible team at Northeastern University's Visualization Lab at Khoury for their feedback. This work was supported in part by grants from the National Science Foundation (Awards \#2236644 and \#2238175).

\bibliographystyle{IEEEtran}
\bibliography{paper}

\newpage
\section{Biography Section}

\vspace{-140mm}
\begin{IEEEbiography}
[{\includegraphics[width=1in,height=1.25in,clip,keepaspectratio]{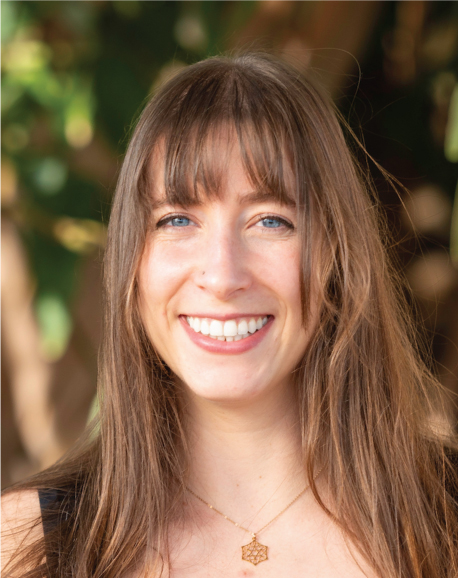}}]{Racquel Fygenson} is a Computer Science PhD Candidate in Northeastern University's Data Visualization Lab. She studies how small changes in visualization design can impact readers' understanding of information.
\end{IEEEbiography}
\vspace{-141mm}
\begin{IEEEbiography}
[{\includegraphics[width=1in,height=1.25in,clip,keepaspectratio]{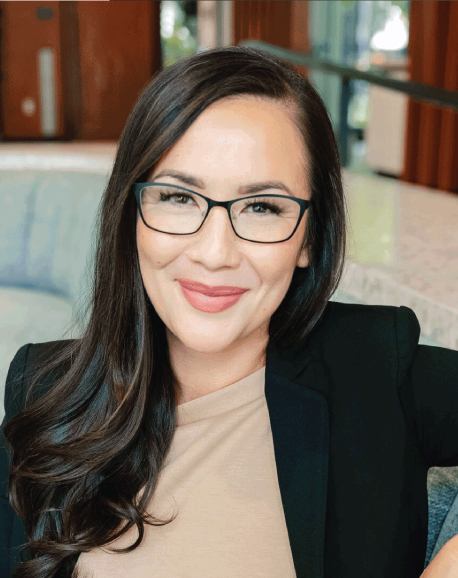}}]{Lace Padilla} is an Assistant Professor at Northeastern University in the Computer Science and Psychology departments. She studies how to align data visualizations with human decision-making capabilities.
\end{IEEEbiography}
\vspace{-142mm}
\begin{IEEEbiography}
[{\includegraphics[width=1in,height=1.25in,clip,keepaspectratio]{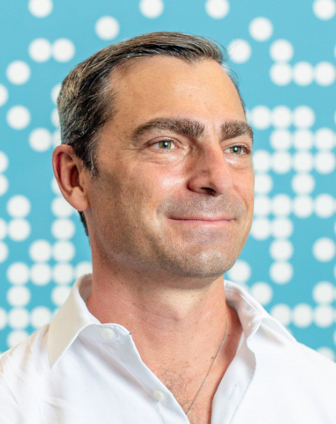}}]{Enrico Bertini} is an Associate Professor at Northeastern University with a double appointment between Computer Science (Khoury) and Art/Media/Design (CAMD). His research focuses on the study of interactive visual interfaces to help scientists, researchers, and domain experts reason with data and models.
\end{IEEEbiography}

\end{document}